\documentclass{ckm}                 

\usepackage{mathptmx}          

\confname{Workshop on the CKM Unitarity Triangle, IPPP Durham, April
  2003}

\title{The LHCb Trigger Strategy and Performance}

\author{M Ferro-Luzzi\addressmark{a}, in behalf of the LHCb Collaboration.} 

\address[a]{CERN, Geneva}

\begin{document}

\begin{abstract}
We give an outline of the LHCb trigger strategy and performance.
The first and second levels (called L0 and L1) are discussed
in some detail, while the subsequent Higher Level Trigger (HLT),
which is currently under development, is only briefly described.
\end{abstract}

\maketitle


\section{Introduction}

\begin{figure}
\hbox to\hsize{\hss
\includegraphics[width=\hsize]{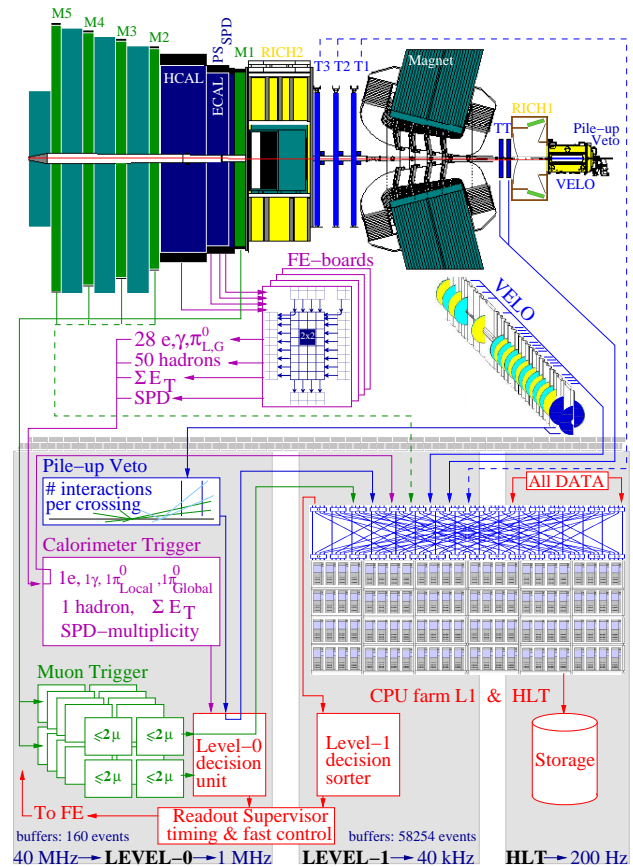}
\hss}
\caption{Overview of the LHCb detector and trigger scheme.}
\label{overview3.eps}
\end{figure}

Precise measurements of CP-violating asymmetries, oscillation parameters 
and branching ratios on numerous B-decay channels will allow an
overdetermination of CKM parameters, with possible inconsistencies pointing 
to physics beyond the Standard Model.
In order to achieve this, an experiment capable of triggering on 
the various decay modes of B-mesons (in particular hadronic modes)
is desirable.
LHCb~\cite{LHCb-TP} is an experiment dedicated to such precision studies
at the large hadron collider (LHC). 
It is designed as a forward spectrometer, because ${\rm B}\bar{\rm B}$ pairs 
are expected to be predominantly 
produced at small polar angles.
An overview of the LHCb detector and trigger scheme is shown in
Figure~\ref{overview3.eps}.

The trigger contains three levels, called Level-0 (L0), Level-1 (L1) and 
Higher Level Trigger (HLT).
\begin{itemize}
\item[\bf L0] uses information from the Pile-up Veto, the
calorimeters and the muon chambers.
All electronics are implemented in full
custom boards, however only commercial components are used.
Part of the functionality of the calorimeter triggers is placed in an 
environment which is expected to receive a few hundred rad per year,
all other hardware is housed in a radiation-free region.
L0 is fully synchronous, {\sl i.e.}
its latency does not depend upon occupancy nor on history. 
The front-end electronics allow a maximum
latency of 4 $\mu$s, and the maximum output rate is limited to 1.1 MHz due
to the multiplexing of the FE electronics of the other sub-systems.
\item[\bf L1] is based on two silicon tracker systems (the vertex locator, 
VELO~\cite{Velo-TDR}, and the trigger tracker, TT) and on the summary 
information of L0.
The trigger algorithm is implemented on a commodity CPU farm.  Its maximum
output rate is 40 kHz, at which rate full event building is performed.
\item[\bf HLT] has access to the full event data,
and is executed on the same commodity CPU farm.
The algorithm first confirms 
the L0 and L1 triggers with better precision, and then will mimic the
off-line selection algorithms for the various channels to reduce 
the rate to 200 Hz,
at which rate events will be written to storage. The HLT algorithms are
under development, and will be described in more detail in the forthcoming
LHCb trigger technical design report.
\end{itemize}

The L0 and L1 triggers have been simulated in the standard LHCb
software framework, which includes an event generator tuned to the LHCb 
phase-space (Pythia 6.205~\cite{Pythia6.2}),
a detailed material description of the detector
(in Geant3) and a realistic description of the digitization stage with 
various sources of background (electronics noise, cross-talk,
machine background, etc.).

\section{Level-0 trigger}
\def\ET{E_{\rm T}}
\def\pT{p_{\rm T}}
The L0 trigger has two distinct components: on the one hand 
B-meson decay products, such as large $\ET$ leptons and hadrons, are
reconstructed, while on the other hand global event variables such as the number
of interactions and multiplicities are collected. The former are
used to distinguish interactions with interesting B-meson decays from the
minimum-bias background, while the latter are used to assure that the
events are selected based on the B signature rather than because of 
large combinatorics.
Thus, it is avoided that these events occupy
a disproportional
fraction of the data-flow bandwidth or available processing power.

The muon chambers allow stand-alone muon reconstruction with a
$\pT$ resolution of $\sim\! 20\%$~\cite{Muon-TDR}. The chambers are 
subdivided into 120k pads and strips.
Pads and strips are combined to form 26k so-called logical pads, 
which range in size from 1.0$\times 2.5$\,cm$^2$ near the beam to 
25$\times$31\,cm$^2$ for the pads in M5 furthest away from the beam.
All pads are projective in the non-bending plane. 
One crate per quarter houses the
trigger boards which reconstruct the two muons with the largest
$\pT$. 
There are no cross connections between the crates, 
and hence muons crossing the quarter boundaries are not reconstructed.

The calorimeter system~\cite{Calo-TDR} provides the following information
for the L0 trigger:
\begin{enumerate}
\item The electromagnetic calorimeter (ECAL) is of the shashlik type, 25
radiation lengths thick, contains 5952 cells, and provides 8-bit $\ET$
information per cell.
\item The preshower (PS)  collects the light after 2.5 radiation lengths of 
lead, is also subdivided in 5952 cells, and provides one bit per cell
for e/$\pi$ separation by setting a threshold that
depends on the radial position of the cell.
\item The scintillating pad detector (SPD) distinguishes between 
charged and neutral 
particles which produce a shower in the ECAL, and
consists of 5952 cells, providing one bit per cell.
\item The hadronic calorimeter (HCAL) is constructed of iron/scintillating 
tiles subdivided into 1468 cells and also provides 8-bit $\ET$ 
information per cell.
\end{enumerate}
The implementation of the calorimeter trigger 
is based on forming clusters by adding the $\ET$ of 2$\times$2 cells, 
and selecting the
clusters with the largest $\ET$. Clusters found in the ECAL are identified
as e,$\gamma$ or hadron depending on the information from the PS and SPD.
The largest HCAL clusters have the energy of the corresponding ECAL cluster
added to them if this ECAL cluster is the largest cluster in an area of
4$\times$8 cells and matches the HCAL cluster position.  By summing all
transverse energy in 4$\times$8 cells in the ECAL so-called local-$\pi^0$
candidates are formed. Largest $\ET$ clusters on neighbouring groups of
4$\times$8 cells in the ECAL are combined to form so-called global-$\pi^0$
candidates. The $\ET$ of all HCAL cells is summed to provide global event
information to allow an interaction trigger.  The total number of SPD cells
with a hit are counted to provide a measure of the charged track
multiplicity in the crossing, a parameter that is used to reduce
the data size and processing time in the L1/HLT CPU farm while
preserving the signal efficiencies~\cite{SPDmult}. 

The proposed physics measurements at LHCb are best done with events that
contain only 1 or 2 primary vertices. For this reason the luminosity at LHCb
will be tuned in the range 2--5$\times 10^{32}~{\rm cm}^{-2}{\rm s}^{-1}$.
To further increase the event fraction with a single interaction, the Pile-up 
Veto detector is incorporated in L0. It uses four silicon sensors of the same
type as those used in the VELO to measure the radial position of tracks
and distinguish between crossings with single and multiple visible interactions. 
The sensors are subdivided in two stations located upstream of the
interaction point, covering $-4.2<\eta<-2.9$. For tracks coming from the
beam-line the radial position $r$ of a track passing the two stations at
z$_A$ and z$_B$ is related to their origin by
$z_{\rm vertex} = (r_B z_A - r_A z_B)/(r_B-r_A)$. The sensors provide
2048 binary channels using the Beetle front-end chip~\cite{beetle}.  The
radial hits are projected into an appropriately binned histogram according
to the above relation using FPGAs. 
All hits contributing to the
highest peak in this histogram are masked, after which the height of the
second peak is a measure of the number of tracks coming from a second
interaction in the crossing. 
Apart from the backward track multiplicity in the first and second
vertex found, the Pile-up Veto provides the position of these vertex
candidates along the beam-line and the total hit multiplicity in the two
stations. The Pile-up Veto information allows a relative luminosity
measurement~\cite{Lumi}.

The L0 decision unit (L0DU) collects all information from L0 components to
form the L0 trigger, {\sl i.e.} the largest $\ET$ e, $\gamma$, $\pi^0_{\rm
local}$, $\pi^0_{\rm global}$, and the two largest $\ET$ hadron
clusters. Global event variables are also collected: the SPD multiplicity,
and the sum of the transverse energy of the HCAL of the actual crossing,
and of the two preceding and following crossings.  From the
possible eight muons provided by the four quadrants of the muon trigger the
three largest in $\pT$ are selected.  Finally the Pile-up Veto
information is also used.  The L0DU is able to perform
simple arithmetic to combine all signatures into one decision per
crossing. The algorithm employed at the moment accepts events where at
least one of the largest $\ET$ e, $\gamma$, $\pi^0_{\rm local}$,
$\pi^0_{\rm global}$, hadrons or muons is above the trigger threshold for the
corresponding particle type (about 2.6, 3, 4.8, 4.9, 3.5 and 1.3 GeV, respectively), 
providing the Pile-up Veto detects less than three tracks coming from a second 
primary vertex.  Events are also accepted if the sum of the $\pT$ of the two
muons with the largest transverse momentum are above a threshold,
irrespective of the Pile-up Veto result. Other global event variables like
hit multiplicities of the Pile-up Veto and SPD are not yet used in the
results presented below, but instead are considered as contingency.

The above mentioned thresholds have to be set such that the output rate of 
L0 is 1.1 MHz.
The L0 hadron trigger plays a central role in LHCb, occupying
approximately 60~\% of the total L0 bandwidth while
the muon/di-muon and e/$\gamma$/$\pi^0$ triggers fill each about 
20~\%.
Note also that, of the 40 million nominal LHC bunch crossings per second, 
only 75~\% occur with oppositely moving bunches, and in only about 
1/3 of these cases is a collision expected 
which produces at least 2 tracks in the LHCb detector.
Hence, the ``visible" event rate is about 10 MHz.
Therefore, the LHCb L0 trigger reduces the visible rate by only 
a factor of 9, by applying relatively soft $\ET$ and $\pT$ cuts.
After this modest rate reduction, a full software trigger 
is used to further filter the events.

\def\effLzero{\varepsilon_{\rm L0}}
\def\effLone{\varepsilon_{\rm L1}}
\def\effLboth{\varepsilon_{\rm L0L1}}
\def\Bd{\rm B_d^0}
\def\Bs{\rm B_s^0}
\def\Jpsimm{J/\psi(\mu^+\mu^-)}
\def\Jpsiee{J/\psi({\rm e^+e^-})}
\def\to{\rightarrow}

\section{Level-1 trigger}
The L1 trigger exploits the finite life time of the B-mesons in addition
to the large B-meson mass as a further signature to improve the purity
of the selected events. The following information is used by L1:
\begin{enumerate}
\item The L0DU summary information as described in the previous section.
\item The VELO measurements of
the radial and angular position of the tracks, in silicon planes
perpendicular to the beam-line between radii of 8\,mm and 42\,mm.
The angular position is measured 
with quasi-radial strips with a stereo angle between 10--20$^\circ$.
A cluster search algorithm is performed 
in the 170k channels using FPGAs to
find roughly 1000 clusters per event. 
\item The TT  measurements from its four silicon planes, 
two with vertical strips and two with a $\pm 5^\circ$ stereo angle.
About 400 clusters are found in 144k channels using the same implementation
as the VELO and a similar algorithm.
\end{enumerate}
The L1 trigger algorithm will be executed on $\sim 500$ commodity CPUs,
and requires event building at 4 kbytes/event to be performed at a L0 output
rate $\leq 1.1$ MHz. 

\begin{figure}
\hbox to\hsize{\hss
\includegraphics[width=\hsize]{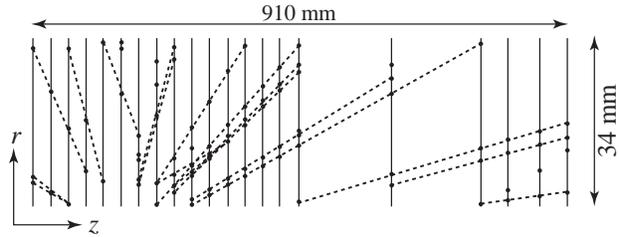}
\hss}
\caption{Event display of the result of the 2D tracking in the VELO 
detector, showing all hits and reconstructed tracks in a slice of 
45$^\circ$ of the VELO $R$-sensors in an event where 72 forward 2D tracks were
reconstructed.}
\label{l1-event}
\end{figure}

B-mesons with their decay products in the LHCb acceptance move
predominantly forward along the beam-line, which implies that the
projection of the impact parameter in the plane defined by the beam-line
and the track is large, while in the plane perpendicular to the beam it is
almost indistinguishable from primary tracks.  The L1 algorithm exploits
this by reconstructing so-called 2D tracks using only the VELO sensors
which measure the radial position.  The 2D track finding efficiency for
charged tracks originating from a B-meson decay and inside the acceptance
of the spectrometer is $\sim 98~\%$.  The 2D tracks are also sufficient to
measure the position of the primary vertex since the strips at constant
radius are segmented in 45$^\circ$ $\phi$-slices.  The RMS of the primary
vertex resolution obtained is 170\,$\mu$m and 50\,$\mu$m in the directions
along and transverse to the beam respectively.  Figure~\ref{l1-event} shows
an event display of the result of the 2D track search in a 45$^\circ$ slice
of the VELO. In this event 72 forward tracks are found in total, while the
mean number of forward tracks in L1 events is 58.

\begin{figure}
\hbox to\hsize{\hss
\includegraphics[width=\hsize]{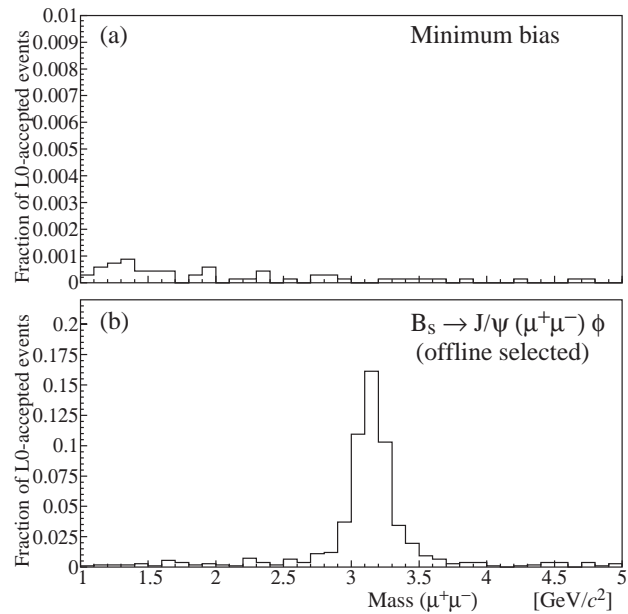}
\hss}
\caption{Invariant mass of $\rm J/\psi$ candidates, for which the muons 
have been reconstructed by combining VELO tracks with the L0 muon information.}
\label{dimu}
\end{figure}

Half of the clusters in the VELO are used to measure the impact
parameter of tracks, and make a preselection on B-decay candidates.  Using
the sensors measuring the angular position, the candidate tracks with an
impact parameter between 200\,$\mu$m and 3\,mm are then converted to tracks
in three dimensions (3D tracks).  The B-decay candidates are also matched
to the electron and hadron clusters from the L0-calorimeter trigger, and
the L0-muon candidates. In Fig.~\ref{dimu} the invariant mass formed from
all oppositely-charged pairs of L0-muon candidates that have been matched
to 3D VELO tracks is shown. A clear $\rm J/\psi$ signal can be
seen, while in minimum bias events only a small fraction of the events
have a $\mu\mu$-candidate with an invariant mass above 2 GeV. 

About eight 3D tracks per event are selected based on their
impact parameter and matched to hits in TT
to measure their momenta.  The 3D VELO tracks are considered matched if
at least three hits are found in the four TT planes.
The magnetic field distribution allows a momentum resolution of
about 20--40$\%$ depending on the momentum, which is sufficient to use the
$\pT$ of tracks as a B signature, 
and also allows
the error on the impact parameter to be calculated including multiple
scattering. The field in between the VELO and TT is parametrized to take its
non-uniformity into account.

\begin{figure}
\hbox to\hsize{\hss
\includegraphics[width=\hsize]{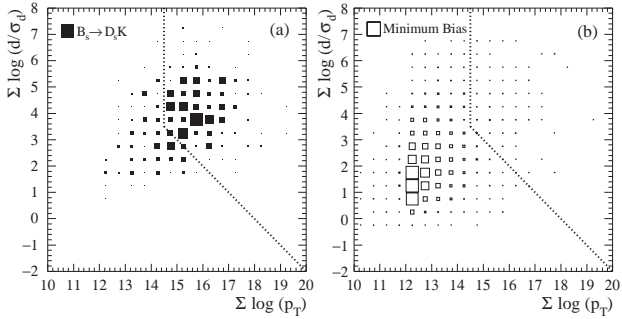}
\hss}
\caption{The logarithmic sum of the $\pT$ of the two VELO--TT tracks 
with the largest $\pT$ in the event versus the logarithmic sum of their
impact parameter significance for (a)~off-line selected $\rm
B_s^0\rightarrow\rm D_s^-\rm K^+$ events, and (b)~minimum-bias events
which have been accepted by L0.  Indicated is the cut which
selects 4$\%$ of the minimum-bias events.}
\label{bpipi}
\end{figure}

The final L1 trigger decision is made by combining the information for
tracks with significant impact parameter, large $\pT$ and possibly being
matched to leptons and hadrons from L0.  Figure~\ref{bpipi} shows how the
$\pT$ and impact parameter significance are used to distinguish between the
minimum-bias background events and, in this example, the channel 
$\rm B^0_s\rightarrow D_s^\mp K^\pm$.
Events are selected according to the
logarithmic sum of the $p_{\rm T}$ of the two VELO--TT tracks with the
largest $p_{\rm T}$ in the event, and the logarithmic sum of their impact
parameter significance.  Events are also
accepted if the invariant mass formed from two oppositely-charged pairs of
L0-muon candidates exceeds 2\,GeV$/c^2$.
Channels with leptons in the final state especially profit from the
matching between VELO tracks and L0-lepton candidates, while the VELO--TT
and VELO--L0-hadron matching boosts the efficiency for hadronic final
states compared to just exploiting the impact parameter information from
the VELO.

\section{Summary}

We have presented the LHCb trigger strategy and performance.
Table~\ref{tab-l0l1eff} shows the efficiency of the L0, L1 
and combined L0$\times$L1 triggers relative to off-line selected 
events for a sample of physics channels.
The last column gives the expected annual yield of untagged B decays
useful for physics analysis.
Studies are now concentrating on developing the HLT with efficiencies 
in excess of 90~\% and on optimising the trigger chain with respect
to {\sl flavour-tagged} physics events.
\begin{table}
\begin{center}
\begin{tabular}{|l|r|r|r|r|}
\hline
Channel              & $\effLzero $ & $\effLone $ & $\effLboth $ & Yield \\[2mm]
\hline\hline
$\Bd \to \pi^+\pi^-$ &   61~\%      &  51~\%      &   31~\% & 27 k   \\[2mm]
$\Bd \to  K^+ \pi^-$ &   60~\%      &  49~\%      &   29~\% &115 k   \\[2mm]
$\Bs \to  K^+   K^-$ &   57~\%      &  48~\%      &   27~\% & 35 k   \\[2mm]
\hline
$\Bs \to D_s^-\pi^+$ &   46~\%      &  53~\%      &   24~\% & 72 k   \\[2mm]
$\Bs \to D_s^+  K^-$ &   44~\%      &  65~\%      &   29~\% &  8 k   \\[2mm]
\hline
$\Bs\to \Jpsimm\phi$ &   93~\%      &  73~\%      &   68~\% &109 k   \\[2mm]
$\Bs\to \Jpsiee\phi$ &   52~\%      &  43~\%      &   22~\% & 19 k   \\[2mm]
\hline
$\Bs\to\Jpsimm K^0_s$&   91~\%      &  71~\%      &   65~\% &119 k   \\[2mm]
$\Bs\to K^{*0}\gamma$&   82~\%      &  33~\%      &   27~\% & 20 k   \\[2mm]
\hline
\end{tabular}
\caption{Efficiencies of the L0, L1 and combined L0$\times$L1 triggers 
relative to off-line selected and untagged events. The last column gives
the expected annual yield of interesting B decays.}
\label{tab-l0l1eff}
\end{center}
\end{table}

\end{document}